\documentclass{osa-article}

\journal{osac}

\articletype{Research Article}

\begin{document}

\title{Generative adversarial network based single pixel imaging}

\author{Ming Zhao,\authormark{1,*} Fengqiang Li, \authormark{2} Fengyue Huo, \authormark{1} Zhiming Tian \authormark{1}}

\address{\authormark{1}College of Information Science Technology, Dalian Maritime University, Dalian, 116026, China\\
\authormark{2}Department of Computer Science, Northwestern University, Evanston, IL, 60208, USA}

\email{\authormark{*}eachzhao@126.com} 



\begin{abstract}
Single pixel imaging can reconstruct two-dimensional images of a scene with only a single-pixel detector. It has been widely used for imaging in non-visible bandwidth (e.g., near-infrared and X-ray) where focal-plane array sensors are challenging to be manufactured. In this paper, we propose a generative adversarial network based reconstruction algorithm for single pixel imaging, which demonstrates efficient reconstruction in 10$m$s and higher quality. We verify the proposed method with both synthetic and real-world experiments, and demonstrate a good quality of reconstruction of a real-world plaster using a 5$\%$ sampling rate.
\end{abstract}

\section{Introduction}
\label{sec:intro}
Single pixel imaging (SPI) is an imaging technique which uses only a single pixel detector to record the reflection from a scene, and computationally generates a two-dimensional (2D) reconstruction of the scene~\cite{duarte2008single}. This method has been widely used for imaging in non-visible bandwidth where focal-plane array sensors are expensive or challenging to be manufactured. For example, it has been used for X-ray imaging~\cite{greenberg2014compressive,yu2016fourier}, near-infrared imaging~\cite{radwell2014single}, and THz imaging~\cite{watts2014terahertz}. SPI has also been extended for ghost imaging in quantum physics~\cite{shapiro2008computational} and three-dimensional imaging~\cite{sun2016single,li2017cs}. 

To reconstruct a 2D scene with single pixel detector, we spatially modulate the illumination or modulate the reflection from the scene before collecting by the detector~\cite{duarte2008single}. A spatial light modulator (SLM) such as digital mirror device (DMD) is usually used to modulate the illumination or reflection from the scene. Random or Hadamard patterns can be displayed on the SLM for post reconstruction~\cite{duarte2008single}. Multiple single-pixel detector measurements with varying modulation patterns (a.k.a sampling mask) are recorded to reconstruct the 2D scene. 

Reconstruction of a 2D scene from single pixel measurements is an ill-pose problem where the reconstruction is not unique~\cite{candes2006stable}. It has been formed as an optimization problem to generate an optimal 2D reconstruction of the scene (see more details in Sec.~\ref{single_pixel_recon}). Compressive sensing~\cite{donoho2006compressed} has been used for reconstruction with different regularizers such as total variation~\cite{duarte2008single,4286571,wakin2006architecture}. Metzler \textit{et al.}~\cite{metzler2016denoising} use approximate message passing (AMP) framework and consider denoising as part of the reconstruction to produce a higher-quality reconstruction (named as D-AMP).

Recently, deep learning~\cite{lecun2015deep} has been used for single pixel imaging or compressive sensing reconstructions. Kulkarni \textit{et al.}~\cite{kulkarni2016reconnet} take convolutional neural network (CNN) as part of the compressive sensing reconstruction, named as ReconNet, which produces better performance compared to iterative optimization methods. Similarly, Higham \textit{et al.}~\cite{higham2018deep} use a three-layer CNN for single pixel imaging which has comparable reconstructions compared to previous methods. Yao \textit{et al.}~\cite{yao2019dr2} use residual learning blocks and linear mapping to further improve the quality of learning based compressive sensing reconstruction, which is named as DR$^2$-Net. Li \textit{et al.}~\cite{li2020compressive} also take deep learning into ghost imaging through scattering media with single pixel detector.
 
Following the learning based reconstruction, we propose a generative adversarial network based SPI, which provides better reconstruction compared to previous learning based methods~\cite{kulkarni2016reconnet,yao2019dr2}. Optimal sampling masks on the SLM are also learned through the proposed neural network. We verify the proposed method with both synthetic and real-world experiments, and different sampling rates of 0.1, 0.05, and 0.01 are used for the evaluation. The rest of the paper is organized as follows: In Sec. 2, we introduce the SPI acquisition system and details of the proposed method. In Sec. 3, we verify our method with both synthetic and real-world experiments. In Sec.4, we discuss the reconstruction with RGB channels and ideas of robust reconstruction and reconstruction with even lower sampling rate with Hadamard patterns. Finally, Sec. 5 concludes the paper.

\begin{figure}[htp]
	\centering
\includegraphics[width=0.9\linewidth]{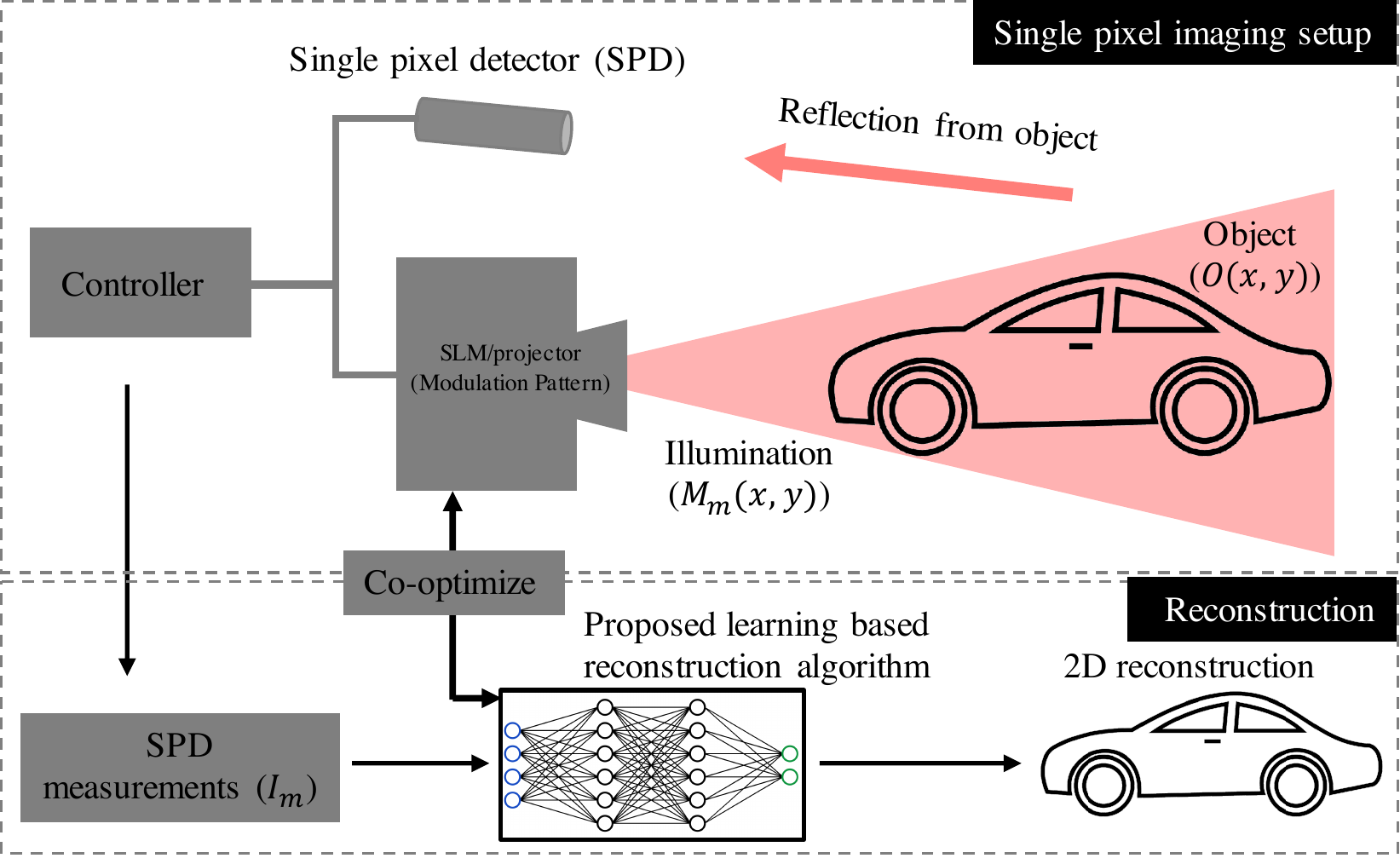}
\caption{\textbf{Overview of the system:} In single pixel imaging, a SLM (projector) with sampling mask displayed on is used to modulate the illumination beam in this work. The beam then hits the object and the reflection is collected by a single pixel detector (SPD). We propose a learning based method for reconstruction from SPD measurements and also learn sampling masks displayed on the SLM.}
\label{fig:highlevel}
\end{figure}

\section{Generative adversarial network based single pixel imaging}
\label{sec:format}
\subsection{Single pixel imaging}
\label{single_pixel_recon}
In single pixel imaging, the illumination is first spatially encoded with a sampling mask $M_m(x,y)$ displayed on an SLM (projector), and then illuminates the scene as shown in Fig.~\ref{fig:highlevel}. The reflection from the object $O(x,y)$ is collected by a single pixel detector (SPD), and the SPD readout $I_m$ can be represented as below:
\begin{equation}
\small
    I_m = \int_{0}^{\Delta t}\int_{0}^{N_y}\int_{0}^{N_x}M_m(x,y)\cdot O(x,y,t) dxdydt+n_m
\end{equation}
where $\Delta t$ is the detector's exposure time. $N_x$ and $N_y$ are the number of pixels in x and y axis. $n_m$ is the measurement noise. In this work, we focus on static scene.

Assume the measurement matrix of the whole system as $\textbf{A}$ (encoding the sampling mask), we then simplify the measurement $I$ as:
\begin{equation}
    \mathbf{I} = \textbf{A}\mathbf{O}
\label{eq:eq2}
\end{equation}
where $\mathbf{I}$ $\in$ $\mathbb{R}^{J\times 1}$, $\mathbf{A} $ $\in$ $\mathbb{R}^{J\times K}$ ($J \leq K$), $\mathbf{O} $ $\in$ $\mathbb{R}^{K\times 1}$. $J$ is the number of measurements with different sampling masks on the SLM, and $K$ is the size of the vector form of the reconstruction ($K = N_x \times N_y$). The \textit{sampling rate} (SR) is defined the ratio of $J/K$, and low SR leads to less number of measurements.

The reconstruction from SPD measurements has traditionally been formulated as an optimization problem with a regularizer for the penalty.
\begin{equation}
    \mathbf{O} = \arg\!\min||\mathbf{I}-\textbf{A}\mathbf{O}||_2^2 + \lambda\Phi (\mathbf{O})
\end{equation}
where $\Phi(\mathbf{O})$ is a regularizer and $\lambda$ is the weight. As shown in Fig.~\ref{fig:highlevel}, we propose a learning based method for the reconstruction from SPD measurements.

\begin{figure}[htp]
	\centering
\includegraphics[width=1\linewidth]{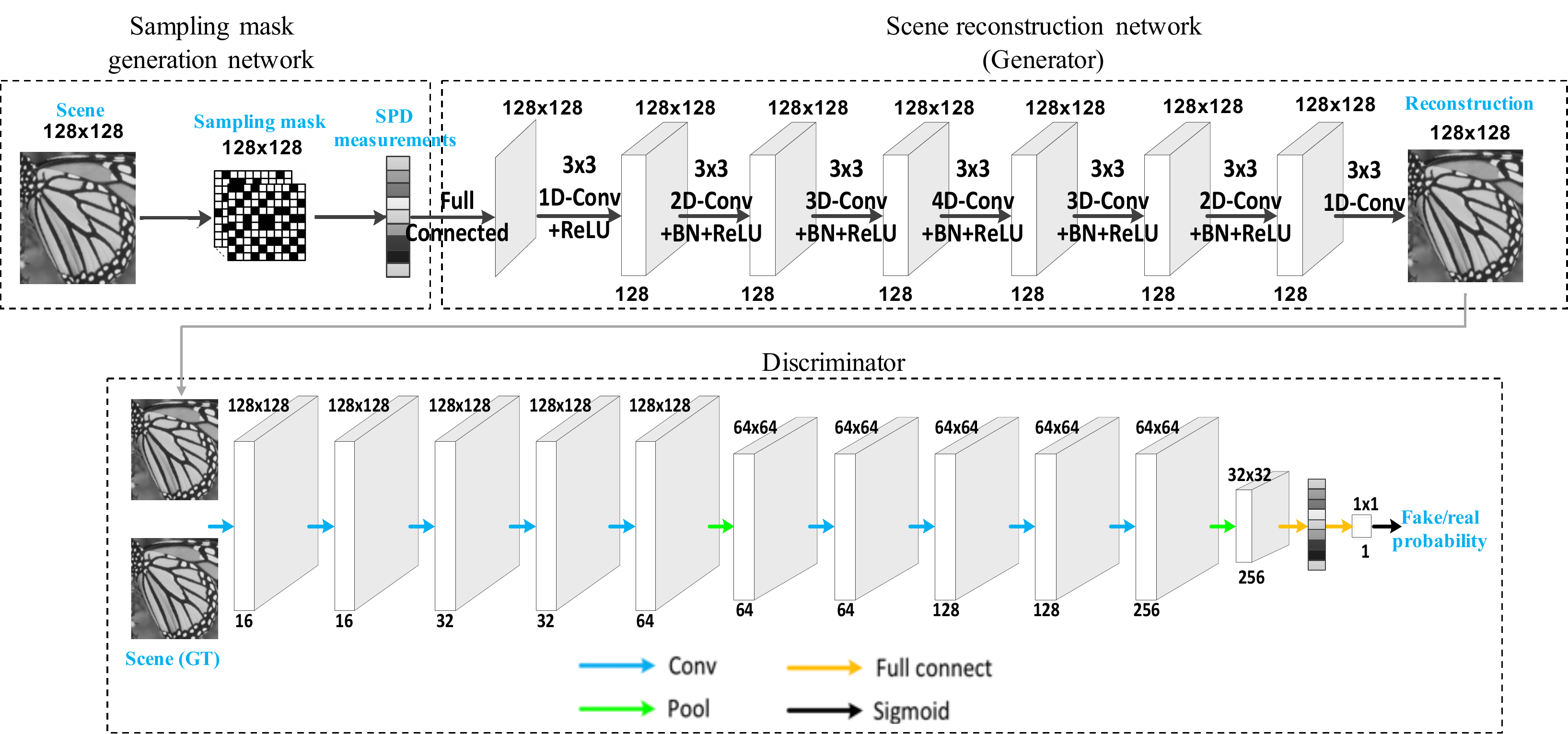}
\caption{The proposed neural network: In training procedure, it contains three parts: sampling mask generation, scene reconstruction (generator), and adversarial network (discriminator). For inference, we use generator to reconstruct the scene with SPD measurements using trained models. Both $W$ and $H$ are 128 in this work.}
\label{fig:nn}
\end{figure}

\subsection{Proposed learning based method}
We propose a generative adversarial network (GAN) based single pixel imaging as shown in Fig.~\ref{fig:nn}. Instead of using random or Hadamard sampling mask, we use a learned sampling mask with the training dataset. In training procedure, it contains three parts: sampling mask generation, scene reconstruction, and adversarial network. 

The sampling mask is learned with the neural network, and each pixel in the mask is -1 or 1. We use binary pattern on SLM for efficiency. In experiment, the measurement with (-1, 1) pattern can be acquired by subtracting the measurement with (0, 1) pattern from that with (1, 0) pattern. As shown in Eq.~\ref{eq:eq2}, if the size of reconstruction is $W \times H$ and the number of measurements is $M$, the measurement matrix size is $M \times (WH)$. We use a fully connected layer to learn these $M$ sampling masks, and there are $M \times W \times H$ parameters with value of 1 or -1. We use the sign function to determine each parameter $\omega$. The sampling mask is learned for each sampling rate.
\begin{equation}
    \omega=
    \begin{cases}
      1, & \text{if}\ \omega>=0 \\
      -1, & \text{otherwise}
    \end{cases}
\end{equation}

We simulate SPD measurements by pixel-wise multiplication of scene and $M$ sampling masks as the input vector $I$ to the scene reconstruction network. There are seven layers for the reconstruction network as shown in Fig.~\ref{fig:nn}. The SPD measurements ($M\times 1$) first go through a fully connected neural network to generate an initial reconstruction of the scene. Then, the initial reconstruction goes through a neural network consisting of five neural layers. In each layer, it contains a batch normalization (BN) layer and the activation layer is rectified linear unit (ReLU). The kernel size in each layer is $3 \times 3$. Finally, a convolution layer is used to generate the reconstruction of the scene. 

For the adversarial network, the input is the reconstruction from the generative network and the ground truth. There are nine convolutional layers. In each convolutional layer, the kernel size is $3 \times 3$, and it contains a BN layer and the activation layer is Leaky ReLU. There are two max pooling layers after the fifth and ninth convolutional layers. For the last two layers, there are fully connected layers and sigmoid function is used as the function layer. 

The loss function ($\mathcal{L}$) contains content and adversarial losses ($\mathcal{L}_{adv}$). The content loss includes pixel-wise MSE loss ($\mathcal{L}_{MSE}$) and VGG loss ($\mathcal{L}_{VGG}$). The VGG loss is based on the ReLU activation layers of the pre-trained VGG19 network~\cite{simonyan2014very}, which is defined as the euclidean distance between the feature representation of the reconstructed scene $G(I, \theta_G)$ and that of reference image $O$.

\begin{equation}
    \mathcal{L} = \mathcal{L}_{MSE} + \mathcal{L}_{VGG} + \lambda \mathcal{L}_{adv}
\end{equation}

\begin{equation}
\small
    \mathcal{L}_{MSE} = \frac{1}{W H}\sum\limits_{x=1}^{W}\sum\limits_{y=1}^{H} \left (O_{x,y} - G(I, \theta_G)_{x,y} \right)^2
\end{equation}
\begin{equation}
\small
    \mathcal{L}_{VGG} = \frac{1}{W_{i,j} H_{i,j}}\sum\limits_{x=1}^{W_{i,j}}\sum\limits_{y=1}^{H_{i,j}} \left (\phi_{i,j}(O)_{x,y} - \phi_{i,j}(G(I, \theta_G))_{x,y} \right)^2
\end{equation}
\begin{equation}
\small
    \mathcal{L}_{adv} = \frac{1}{2} \cdot \left (D(G(I,, \theta_G), \theta_D) - 1 \right)^2
\end{equation}
where $W$ and $H$ represent width and height sizes of reconstruction for the scene ($O$). $I$ is a vector of SPD measurements. $\phi_{i,j}$ represents the feature map of the j-th convolution before the i-th layer in VGG-19 network. $W_{i,j}$ and $H_{i,j}$ are dimensions of feature maps within VGG-19 network. $\theta_G$ and $\theta_D$ are the network parameters in the generative and adversarial networks. $\lambda$ is 0.05.

\subsection{Network training}
We use STL-10 dataset~\cite{coates2011analysis} for the neural network training. We preprocess the size of images in STL-10 dataset to be 128 $\times$ 128, and convert each RGB image into gray scale with value from 0 to 1. We set the ratio of training and validation images to be 9:1. As mentioned previously, we simulate SPD measurements by pixel-wise multiplication of scene and the sampling mask during training.

The learning rate is set to 10$^{-5}$ for the sampling network, 10$^{-4}$ for the reconstruction network, and 10$^{-5}$ for the adversarial network. We use Adam~\cite{kingma2014adam} as the optimization function with $\beta_1$ as 0.9 and $\beta_2$ as 0.99. During training, the adversarial network is updated after four epochs in the reconstruction network. An NVIDIA GTX 2080Ti GPU is used for the training and the framework is implemented with Tensorflow. The training time takes about seventy hours.

After training, the learned model is used for inference with synthetic data and real-world SPD measurements.

\begin{figure}[htp]
	\centering
\includegraphics[width=0.8\linewidth]{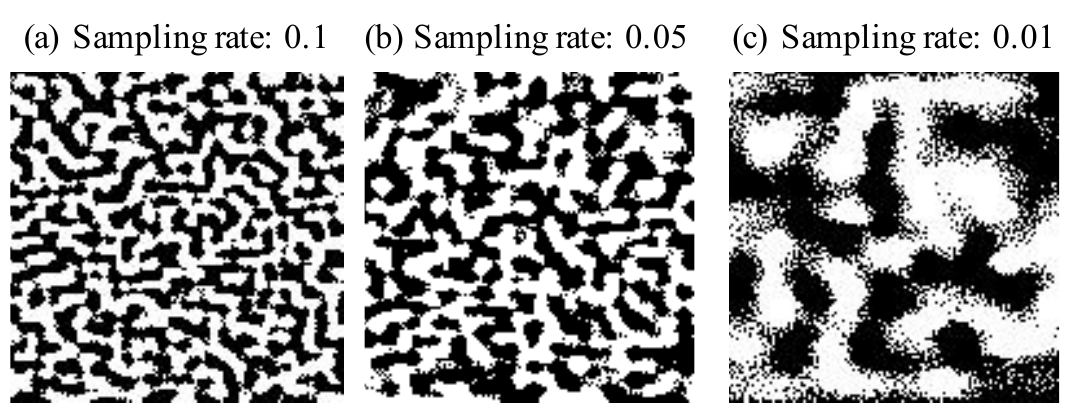}
\caption{(a-c) show one learned sampling mask for SRs of 0.1, 0.05, and 0.01 respectively.}
\label{fig:samplemask}
\end{figure}

\subsection{Sampling mask}
\label{sample_mask}
We trained the neural network individually for different SRs of 0.1, 0.05, and 0.01. Examples of the learned sampling mask for SRs are shown in Fig.~\ref{fig:samplemask} (a-c) respectively. The white and dark areas represent 1 and -1 in the sampling mask. 

As we can see from Fig.~\ref{fig:samplemask}, the high-SR sampling mask has higher frequency feature compared to low-SR sampling masks since higher sampling rates can generally help produce a reconstruction with more fine structures.  We use these learned sampling masks to display on the SLM and modulate the illumination in the real-world experiment in Sec.~\ref{sec:exp}.

\section{Experiments}
\label{sec:simu}
\subsection{Synthetic experiment}
\begin{figure}[htp]
	\centering
\includegraphics[width=0.9\linewidth]{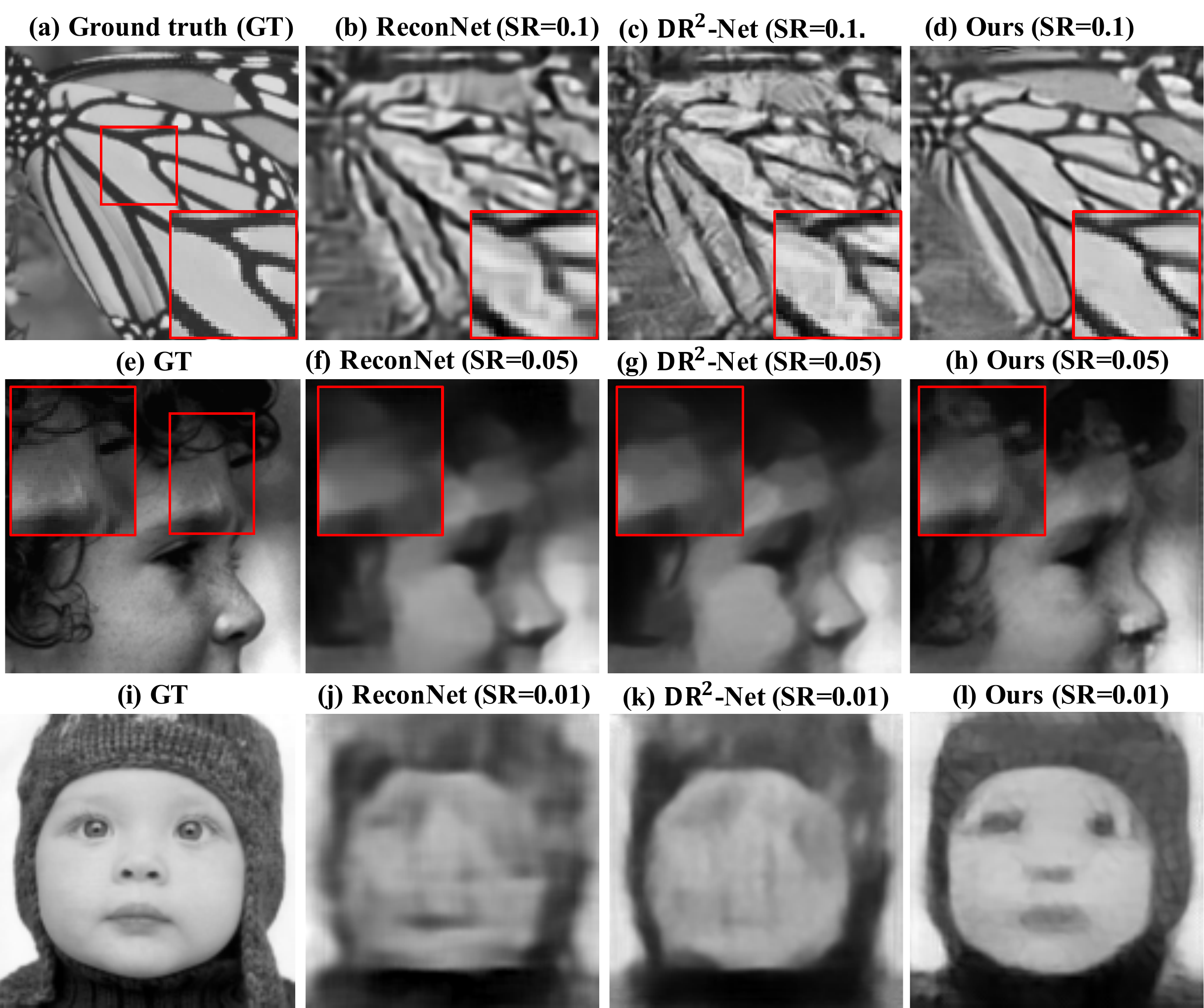}
\caption{Comparison with learning methods: We show reconstructions with ReconNet, DR$^2$-Net and ours for SR = 0.1 (b-c), SR=0.05 (f-h), and SR=0.01 (j-l), respectively. (a, e, i) show corresponding ground truth images.}
\label{fig:simu_res}
\end{figure}

\begin{table}[htp]
\centering
\begin{tabular}{c | c | c | c | c | c | c  } \hline \hline
 & \multicolumn{6}{c}{Set5}\\ \cline{2-7}
 & \multicolumn{2}{c|}{SR=0.1} & \multicolumn{2}{c|}{SR=0.05} & \multicolumn{2}{c}{SR=0.01}  \\ \cline{2-7}
 & PSNR (dB) & SSIM & PSNR (dB) & SSIM  & PSNR (dB) & SSIM  \\ \hline
D-AMP~\cite{metzler2016denoising} & 22.20 & 0.3941	& 17.75	& 0.2130 & 8.51	& 0.0146  \\ \hline
ReconNet~\cite{kulkarni2016reconnet} & 24.37 & 0.7193 & 22.09 & 0.6086 & 17.68 & 0.3847  \\ \hline
DR$^2$-Net~\cite{yao2019dr2} & 24.84	& 0.7434 & 22.31 & 0.6262 & 17.73 & 0.3859  \\ \hline
Ours & \textbf{26.30} & \textbf{0.8130} & \textbf{24.31} & \textbf{0.7098} & \textbf{20.10} & \textbf{0.4841} 
 \\ \hline \hline
 
 &  \multicolumn{6}{c}{BSD100} \\ \cline{2-7}
 & \multicolumn{2}{c|}{SR=0.1} & \multicolumn{2}{c|}{SR=0.05} & \multicolumn{2}{c}{SR=0.01} \\ \cline{2-7}
 & PSNR (dB) & SSIM & PSNR (dB) & SSIM  & PSNR (dB) & SSIM  \\ \hline
D-AMP~\cite{metzler2016denoising} & 23.16 & 0.2954 & 18.96 & 0.1475 & 10.05 & 0.0200 \\ \hline
ReconNet~\cite{kulkarni2016reconnet} & 24.96	& 0.6542 & 23.23 & 0.5702 & 19.72 & 0.4168 \\ \hline
DR$^2$-Net~\cite{yao2019dr2} & 25.46 & 0.6832 & 23.55 & 0.5914 & 19.66 & 0.4155 \\ \hline
Ours  & \textbf{26.56} & \textbf{0.7452}	& \textbf{25.26} & \textbf{0.6684} & \textbf{21.91} & \textbf{0.4734}
 \\ \hline \hline
 
\end{tabular}
\caption{We test on Set5 and BSD100 datasets in the synthetic experiment. We calculate the averaged PSNR and SSIM values with each dataset for different reconstruction methods.}
\label{table:simu}
\end{table}

We first perform a synthetic experiment and test the trained model on Set5~\cite{bevilacqua2012low} and BSD100~\cite{martin2001database} datasets. To acquire the synthetic SPD measurement with different SR of 0.1, 0.05, and 0.01, we perform a pixel-wise multiplication between the learned sampling mask and the reference image and then sum into a single value. We compare our reconstruction results with D-AMP~\cite{metzler2016denoising}, ReconNet~\cite{kulkarni2016reconnet}, and DR$^2$-Net~\cite{yao2019dr2} as shown in Tab.~\ref{table:simu}. For ReconNet and DR$^2$-Net, we use Gaussian random pattern as the modulation patterns in this comparison and the following real-world experimental comparison. Compared to previous methods, our proposed method achieves better performances in both PSNR and SSIM for all sampling rates. Compared to iterative optimization method, learning based methods demonstrate efficient reconstructions. For example, our method can reconstruct a scene within 10$m$s.

We also show examples of reconstructions with different learning methods in Fig.~\ref{fig:simu_res}. As we can see, our proposed method can provide better visual quality with more details of the scene compared to other methods.

\subsection{Real-world experiment}
\label{sec:exp}
\begin{figure}[htp]
	\centering
\includegraphics[width=0.9\linewidth]{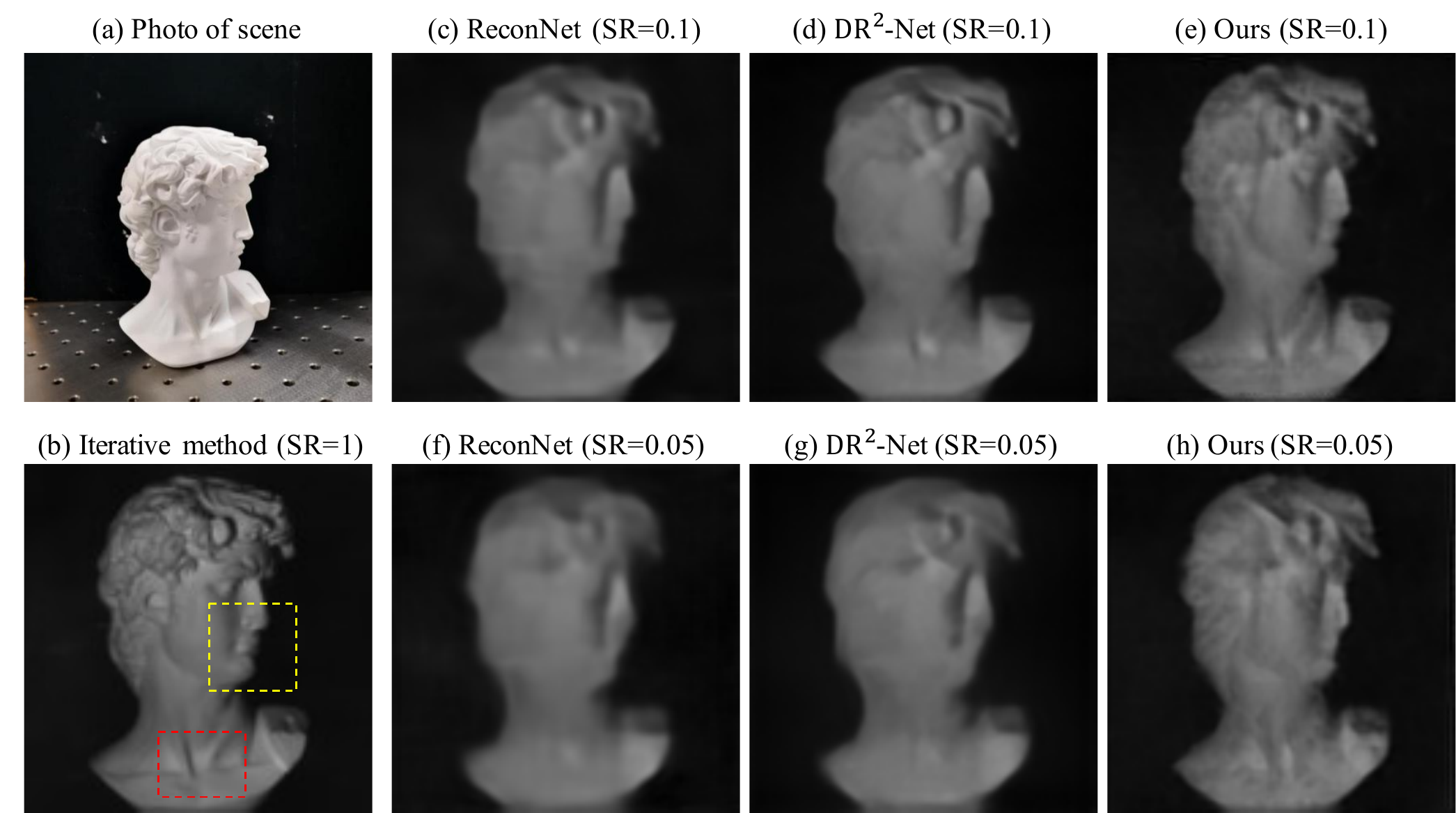}
\caption{We verify the proposed method with a real-world experiment, and a plaster bust (a) is used as the target. (b) shows the reconstruction of iterative optimization method using Hadamard pattern with SR=1. (c-e) show reconstructions with SR=0.1 for different learning based methods. (f-h) show reconstructions with SR=0.05 for learning based methods.}
\label{fig:real_res}
\end{figure}

An experimental system as shown in Fig.~\ref{fig:highlevel} is built to verify the proposed method. A light-emitting diode is used as the illumination source, and the illumination is modulated with sampling mask displayed on a VIALUX V7000 DMD (projector). We use the learned sampling masks from Sec.~\ref{sample_mask} to display on the DMD. A plaster bust (Fig.~\ref{fig:real_res}(a)) is used as the object, and the distance from the object to the detector is about 0.5m. The reflection from the object is collected by a single-pixel photomultiplier tube (Hamamatsu H10493-012).

We first use Walsh–Hadamard pattern with a sample rate of 1 to reconstruct the scene using iterative optimization method, and this reconstruction shown in Fig.~\ref{fig:real_res}(b) is used as the baseline for post comparison. We then reconstruct the scene with ReconNet, DR$^2$-Net, and our method. We test with two different SRs of 0.1 and 0.05 for the learning based methods as shown in Fig.~\ref{fig:real_res}(c-e) and (f-h) respectively. As we can see, our method generates more fine details of the scene such as `mouth' and `neck' regions where the other two learning methods fail to provide. We also quantify RMSE and SSIM values by comparing the learning based reconstructions with the baseline (Fig.~\ref{fig:real_res}(b)). The quantitative result is summarized in Tab.~\ref{table:real}, where shows our method achieves better performance. 

\begin{table}[htp]
\centering
\begin{tabular}{c | c | c | c | c  } \hline \hline
 & \multicolumn{2}{c|}{SR=0.1} & \multicolumn{2}{c}{SR=0.05} \\ \cline{2-5}
 & PSNR (dB) & SSIM & PSNR (dB) & SSIM  \\ \hline
ReconNet~\cite{kulkarni2016reconnet} &28.55	&0.6818	&26.83	&0.6354 \\ \hline
DR$^2$-Net~\cite{yao2019dr2} & 28.96	&0.7445	&27.20	&0.6939 \\ \hline
Ours &\textbf{29.61}	&\textbf{0.7527}	&\textbf{28.67}	&\textbf{0.7111}
 \\ \hline \hline
\end{tabular}
\caption{We test the real object and quantify PSNR and SSIM values for different learning methods compared to iterative reconstruction using Hadamard patterns with SR=1.}
\label{table:real}
\vspace{-0.2in}
\end{table}

\section{Discussion}
There exists an interest in achieving RGB images with single-pixel imaging method. This might be realized with lighting the object at three different wavelength bands in red, green, and blue (e.g., laser diodes emission in red, green and blue wavelength bands). We explore the possibility of reconstructing a RGB image from single-pixel imaging with a synthetic experiment. We use the same dataset for the training. We slightly modified the pipeline for RGB by simulating SPD measurements in each channel and concatenating three channels before loading them to scene reconstruction network. 

\begin{figure}[htp]
	\centering
\includegraphics[width=\linewidth]{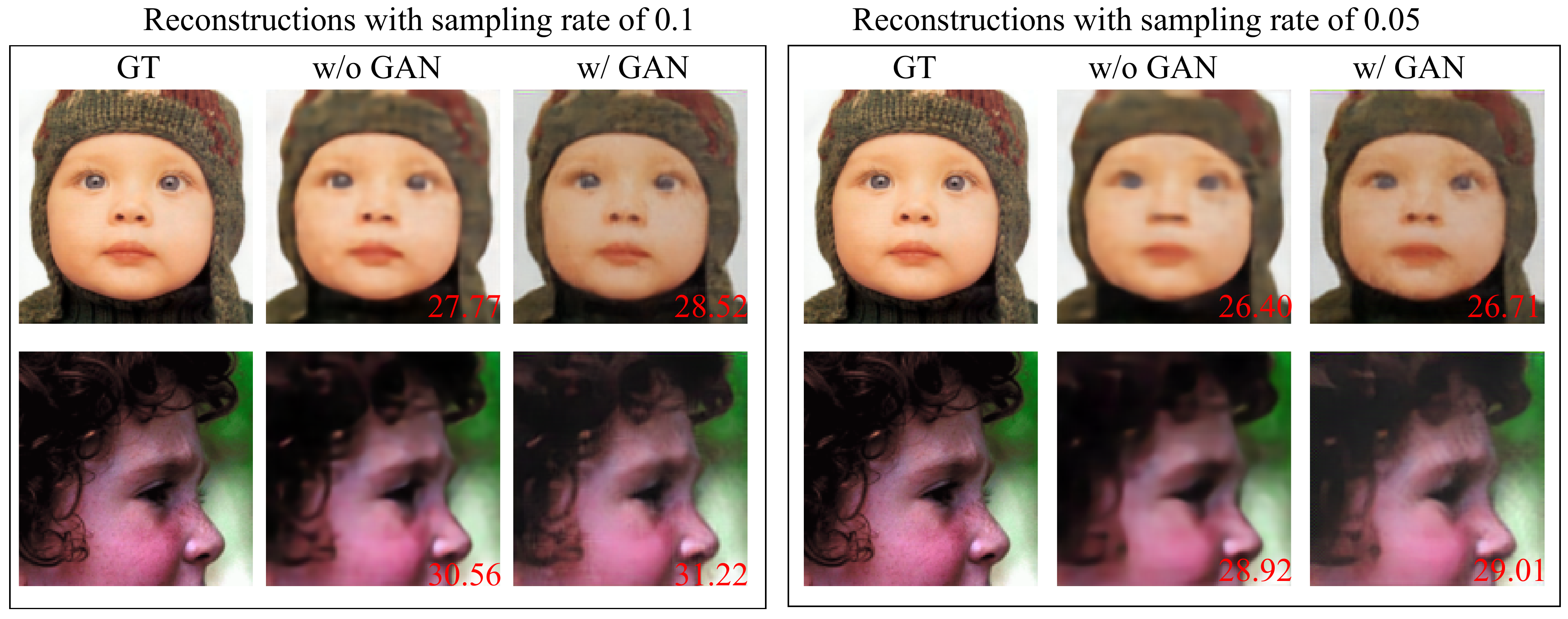}
\caption{We verify the RGB reconstruction with synthetic SPD measurements under different sampling rates of 0.1 and 0.05. We compare the performance with and without GAN in terms of PSNR (marked in red in lower right corner).}
\label{fig:rgb}
\end{figure}

We compare the reconstructions with and without GAN under two different sample rates of 0.1 and 0.05 as shown in Fig.~\ref{fig:rgb}. As we can see the reconstructions with GAN can provide better PSNR performance (marked in red in Fig.~\ref{fig:rgb}) compared to those reconstructions without GAN. Since we concatenate the SPD measurement from three wavelengths bands as input to the reconstruction network, we can turn on red, green and blue lighting source simultaneously in the real-world experiment.

Sun\textit{ et al.} \cite{sun2017russian} demonstrates a "Russian Dolls" idea to rearrange the Hadamard pattern for a high-quality reconstruction with very low sampling rates. Li\textit{ et al.} \cite{li2020compressive} also demonstrates a rearranged Hadamard pattern for robust reconstruction in imaging through scattering media by considering the energy each Hadamard pattern can carry. With the proposed pipeline, we demonstrate the reconstruction with low sampling rate in both synthetic (1\%) and real-world (5\%) experiments. By combing the proposed reconstruction method with these sampling ideas\cite{sun2017russian, li2020compressive} (e.g., using the sampling patterns as initialization for the final learned sampling pattern in our pipeline), we may reconstruct the object with even lower sampling rate to reduce the computation time for imaging processing or have robust reconstruction in complex environment (e.g., imaging through scattering media).

\section{Conclusion}
In this paper, we propose a generative adversarial network based reconstruction algorithm for single pixel imaging to co-optimize the sampling mask and achieve better performance and efficient reconstruction. We verify our method with both synthetic and real-world experiment. With applications of single pixel imaging in X-ray and near-infrared imaging, we believe our method can help higher quality of reconstructions in these applications. We also believe the method can contribute broadly to ghost imaging and imaging scattering media with single pixel detectors.


\bibliography{sample}






\end{document}